\documentclass[aps,prl
,twocolumn
,showpacs
]{revtex4}

\usepackage{epsfig}
\usepackage{bm}
\usepackage{graphicx}
\usepackage[dvips]{color}

\def\be{\begin{equation}}
\def\ee{\end{equation}}
\def\ba{\begin{array}}
\def\ea{\end{array}}
\def\bea{\begin{eqnarray}}
\def\eea{\end{eqnarray}}

\newcommand{\CLR}[1]{{\textcolor[named]{RedViolet}{CLR #1 CLR}}}

\newcommand{\revision}[1]{#1}

\begin{document}

\title{
Plasticity-induced structural anisotropy of silica glass
}

\author{C. L. Rountree$^1$, D. Vandembroucq$^2$, M. Talamali$^2$,
E. Bouchaud$^3$ and S. Roux$^4$}

\affiliation{
$^1$CEA, IRAMIS, SPCSI,~Groupe Fracture \&  Syst\`emes Complexes,
 F-91191 Gif sur Yvette, France\\
$^2$Laboratoire PMMH, UMR 7636 CNRS/ESPCI/Paris 6/Paris 7, 
10 rue Vauquelin F-75231 Paris cedex 05\\
$^3$CEA, IRAMIS, SPEC, Groupe Instabilit\'{e}s \& Turbulence,
 F-91191 Gif sur Yvette, France\\
$^4$LMT-Cachan, ENS de Cachan/CNRS-UMR 8535/Universit\'{e} Paris 6/PRES
    UniverSud Paris\\ 61 avenue du Pr\'{e}sident Wilson, F-94235 Cachan Cedex,
    France}

\begin{abstract}

Amorphous silica density at ambient pressure is known to depend on
thermal history (through the quenching rate) but also, at room
temperature, on the maximum pressure applied in the past. Here we
show that beyond density, a mechanical loading can endow the
structure with an orientational order. Molecular dynamics simulations
show evidence that amorphous silica develops a permanent anisotropic
structure after extended shear plastic flow. \revision{ This
anisotropy which survives for an unstressed specimen  is revealed
markedly by the fabric tensor computed over the Si-O-Si orientations,
albeit the SiO$_4$ tetrahedra microstructure remains mostly
unaltered.}

\end{abstract}

\maketitle

Plasticity of amorphous media, which can be easily evidenced {\it via}
indentation or scratch tests\cite{Taylor-Nat49}, has a very different
nature from its counterpart for crystalline media, since no elementary
entities such as dislocations whose evolution controls plastic flow
can be easily defined \cite{Spaepen-ActaMet77,Argon-ActaMet79}.  The
current view is that spatially distributed local restructuring rather
than extended defect motion (such as dislocation) are responsible for
irreversible strains in amorphous
materials\cite{FalkLanger-PRE98,BulatovArgon94a}.  At a very local
scale, under load, a small group of atoms (called a Transformation
Zone or TZ) may undergo rearrangements, a change of conformation
eventually affecting the topology of the atomic bonds which will
contribute to an elementary increment in irreversible strain.
Although a complete description of these TZ is extremely complex, and
cannot be cast into simple categories, a statistical analysis
capturing the key properties of these zones is an attractive route for
relating the macroscopic mechanical behavior to the underlying
microstructural counterpart\cite{FalkLanger-PRE98}.

In contrast with dislocations which naturally lead to isochoric
plastic deformation, transformation zones may densify or dilate as
they rearrange. Indeed at a macroscopic scale, plasticity of silicate
glasses is known to exhibit permanent densification
\cite{Cohen-PCG65,Ernsberger-JACS68} from a few percents for soda-lime
glasses \cite{Rouxel-ScriptaMat06} to values as large as
20\% in the extreme case of amorphous silica \cite{PMMCVB-JACS06,VDCPBCM-JPCM08}.
This densification naturally affects shear plasticity, and hence
pressure and shear stress are to be coupled in the yield criterion of
amorphous silica~\cite{KBVD-ActaMat08}.

Plasticity of structural glasses is furthermore characterized by a
significant hardening behavior \cite{PMMCVB-JACS06,VDCPBCM-JPCM08}.
The yield surface evolves with the mechanical loading. This means in
particular that, when applying stress in a given direction (pure
shear, pure hydrostatic pressure, etc) the value of the elastic limit
depends on the history of the loading.

To account for this dependence on mechanical history a proper
description of plasticity thus requires the use of additional
internal variables. The first one is obviously density and indeed
a recent study of densification with pressure allows one to
characterize the density hardening of silica~\cite{VDCPBCM-JPCM08}.
The necessity to include more internal variables than the mere density
is a difficult question to address.  Experimentally, plasticity of
amorphous media calls for a high level of stress and confinement which
can either be met in indentation-type test (with the complication of
the very strong spatial heterogeneity of the stress state) or in anvil
diamond cell where shear cannot be imposed.  Thus it appears very
difficult to follow a particular stress path which would allow one to
answer this question.  Hence, in the present study, we will resort to
Molecular Dynamics simulations where homogeneous loadings with
arbitrary stress path can be imposed.

As for granular media plasticity, shear reversal experiments show
that a simple scalar internal variable is not sufficient to account
for the mechanical behavior, and a tensor-valued internal variable,
characterizing the geometry packing is needed \cite{Radjai-inbook04}.
For this purpose, a natural candidate is the so-called fabric tensor,
\revision{$\bm{F}=\langle \bm{n}\otimes \bm{n} \rangle$}, which captures the mean orientation of the
contact normals, \revision{$\bm n$}, through the spatial average of
their dyadic product.
%
%
%
This tensor 
characterizes an anisotropic texture of the medium. Experimental and
numerical studies of granular
media\cite{Oda-MechMat93,Rothenburg-IJSS04}, have  shown that such
was the case after shearing.  Indeed, more contacts are oriented along
the direction(s) of compression. Similarly, the same tools have
recently been used to study the rheology of
foams\cite{Graner-EPJE08}.

The fabric tensor can be characterized by its three eigenvalues,
$\lambda_i$, for $1<i<3$.  The fact that \revision{$\bm n$} is a unit
vector implies that the trace of \revision{$\bm F$} is equal to
unity, or $\sum_i \lambda_i=1$ Thus, for an isotropic
\revision{medium}, the three eigenvalues are equal to 1/3. In case of
anisotropy, degeneracy is lifted and different eigenvalues
$\lambda_i$ are measured. In order to give a quantitative scalar
estimate of this anisotropy, the following scalar $\alpha$ parameter,
proportional to the norm of the deviatoric part of $\bm F$, is
defined:
\begin{equation}
\centering
\alpha = \frac{3}{2} \sqrt{\displaystyle\sum_{i=1}^3{\left({\lambda{_i}
- \frac{1}{3}}\right)^2}}\;.
\label{eq2}
\end{equation}

For an isotropic \revision{medium}, $\alpha = 0$, and the prefactor
has been chosen such that in case of full anisotropy, $\{ \lambda_i
\} = \{1;0;0\}$ we have $\alpha =1$.  \revision{This quantitative
parameter, which is built in the same spirit as an effective shear
stress allows one to compare results obtained in different stress
geometries.} Pursuing our analogy, it is natural to consider a
similar fabric tensor for amorphous silica.  We will show that
indeed, besides the usual (reversible) anisotropy induced by an
elastic strain, plasticity induces a series of structural
rearrangements which after unloading endow silica with a remnant
anisotropic structure.  The latter state can be regarded as a novel
phase of amorphous silica.

\begin{figure}[tb]
\begin{center}
\includegraphics[width=0.45\textwidth]{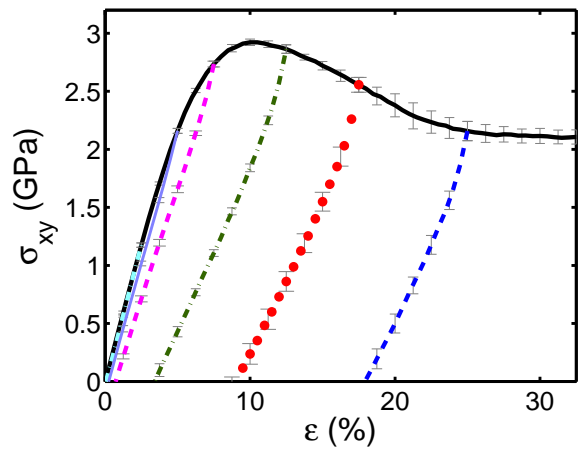}
\caption{ The evolution of the average stress of 12 independent \revision{samples},
composed of 117,912 atoms each, when sheared: (1) from 0 \% to 30 \% (black);
unloaded (2) from +2.5 \%  (sky blue) ; (3) from +5 \% (lavender);
(4) from +7.5 \% (pink); (5) from +12.5 \% (green); (6) from +17.5 \%
(red) and (7) from +25 \% (blue). (Note: the deformation rate is
10$^{-2}$ per ps.)
\label{fig1}}
\end{center}
\end{figure}

\begin{figure}[tb]
\begin{center}
\includegraphics[width=0.48\textwidth]{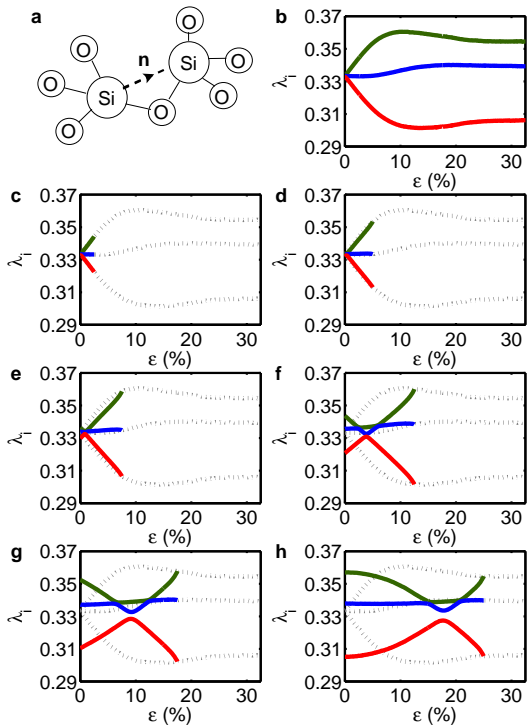}
\end{center}
\caption{Sketch of the ``contacts'' Si-Si between neighbouring SiO$_4$
tetrahedra used to build the fabric tensor \revision{$\bm{F} =\langle
\bm{n}\otimes \bm{n} \rangle$}.
Evolution of the eigenvalues of the fabric tensor when \revision{sheared
from 0 \% to 30 \% (b) and unloaded from: 2.5\% (c); 5\% (d); 7.5\% (e);
12.5\% (f); 17.5\% (g) and 25\% (h).} Before loading the material is
isotropic, the 3 eigenvalues should be equal within the errors bars.
Upon loading (b), degeneracy is lifted and anisotropy first increases
reversibly (c-d) due to the elastic strain. Then plasticity sets in
(e-f-g) and after unloading, the material is left with a remnant
structural anisotropy.
\label{fig2}}
\end{figure}

In the case of silica, chemistry imposes that each silicon atom is
encaged in a tetrahedron of four oxygen atoms forming a SiO$_4$
elementary unit.  These tetrahedra are bounded to each other through
a common oxygen atom. They are extremely robust and remain virtually
unaltered under mechanical loadings. However, pairs of tetrahedra
have much more freedom in their relative orientations.  Henceforth,
it is natural to build the equivalent of a fabric tensor from the
``contacts'' between \revision{neighboring} tetrahedra. Thus for all
Si-O-Si triplets, we consider the unit vector \revision{$\bm n$}
which connects all next to nearest neighbors, NNN, ({\it i.e.} Si-Si
atoms see Fig. \ref{fig2}a). The fabric tensor \revision{$\bm F$} as
defined above is built up by averaging over all NNN
Si atoms. The fact that second neighbors are to be considered is a
challenge to study this anisotropy experimentally using classical
tools for structural analysis.


Amorphous silica was studied via Molecular Dynamics (MD) computer
simulations.
This technique allows us to impose an extended (simple) shear along
one direction under constant volume, and  reverse it.  This
particular stress path is able to reveal whether plasticity is
accompanied by any structural change and in particular anisotropy. MD
simulations are performed on amorphous silica (a-SiO$_2$), using the
empirical interatomic potential developed by Vashishta {\it et
al}~\cite{Vashishta-PRB90,Vashista-NATO97,Rountree-IJF03}. This
potential incorporates steric repulsion, charge transfer, and
electronic polarizability of atoms through pair-wise interaction
terms. Covalent effects in silica are included through bond-bending
and bond-stretching three-body terms. The parameters of the potential
have been adjusted by measurements of structural correlations,
elastic moduli, and fracture
toughness\cite{Vashishta-PRB90,Rountree-ARMR02}. The a-SiO$_2$
\revision{samples} were first prepared by melting (i.e. heating to
4000~K) 117,912 (cubic box of length 12.088~nm) atoms of ideal
$\beta$-cristobalite crystal.  The \revision{sample} was cooled to
2500~K at a rate of 1~K/$\Delta$t (where $\Delta$t is the time step)
and allowed to relax for 60000~$\Delta$t (large \revision{specimen}).
The cooling process was repeated at 1500~K, 600~K, 300~K, and 5~K.
Afterwards a conjugate gradient method is used to \revision{cool the
sample to} 0~K. The temperature is once again elevated to 300~K in
order to conduct simulations at room temperature. \revision{Periodic
boundary conditions have been invoked in all simulations.}

Shear plasticity of this model silica glass has been studied on 12
independent \revision{samples}.  Each sample was sheared by applying
an external shear in small increments such that the volume was
conserved (NVT ensemble).  Between each subsequent shear the sample
was allowed to relax.  Also during the simulations the temperature
was held fixed at 300~K using a thermostat. Two different shear
strain rates have been used, 10$^{-4}$~ps$^{-1}$ and
10$^{-2}$~ps$^{-1}$.  \revision{It is to be stressed that these shear
rates are considerably higher than the highest ones that may be
considered experimentally.  However lower shear rates cannot
conceivably be imposed within the present framework of Molecular
Dynamics. Therefore the effect of long time relaxation processes such
as atom diffusion is out of reach of the present study.}

 Figure \ref{fig1} depicts a typical stress strain curve. The initial
behavior corresponds to linear elasticity; for the faster strain rate
here represented, a stress overshoot is visible before a plateau
corresponding to the stationary plastic regime. As discussed in
\cite{Rottler-PRE03}, the amplitude of this stress overshoot is
dependent on the strain rate and almost disappears for the slower
rate.  In addition to this monotonic test, unloading to zero shear
stress has been performed starting from different values of the
maximum total strain: 2.5\%, 5\%, 7.5\%, 12.5\%, 17.5\% and 25\%.  As
can be seen on Fig. \ref{fig1}, beyond 5\% strain (which corresponds
to a shear stress of about 2.25 GPa), permanent plastic deformation
sets in. \revision{During the simulations the pressure was monitored,
and it was found to increase from $\sim{0GPa}$ during the elastic
loading stage up to $\sim{2.5GPa}$ during plastic flow. Further
details can be found in \cite{Rountree-MDsilica09}.}

In Fig. \ref{fig2} the evolution of the eigenvalues of the fabric
tensor is shown along shear loading and unloading from various values
of total strain. At rest, isotropy of the glass is almost perfectly
obtained; the three eigenvalues are equal within a precision of
$10^{-3}$. Shear loading then induces first an elastic strain which
renders the medium anisotropic.  The elastic shear strain naturally
leads to the lifting of the degeneracy: the eigenvalues take
different values with a growing shift.  In the elastic regime, a
change in fabric tensor can be shown to be proportional to the
strain. Note that we recover here more or less the behavior of the
shear stress: a linear regime related to elasticity followed by a
bump and a stationary elasto-plastic regime. In this regime, the
shear-induced anisotropy is marked and eigenvalues are clearly
separated. \revision{In this plastic flow regime, where the elastic
strain remains invariant, so does the fabric tensor.} When unloading,
the fabric tensor eigenvalues linearly return to their original value
showing that the initial rise of anisotropy is perfectly reversible,
and hence of elastic nature.

\begin{figure}[tb]
\begin{center}
\includegraphics[width=0.45\textwidth]{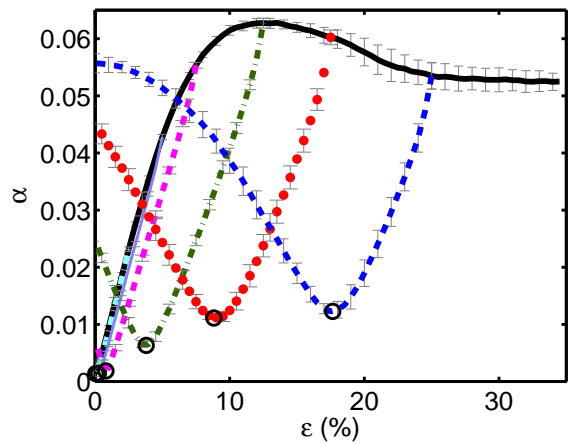}
\end{center}
\caption{
Evolution of the anisotropy index $\alpha$ under shear loading and
unloading (same as above). The minimum value of $\alpha$, black
circles, is associated with the state of the structure under zero
shear stress. Permanent deformation is characterized by a non-zero
value of the anisotropy index: the glass structure has become anisotropic.
\label{fig3}}
\end{figure}

For $\varepsilon_{max}=2.5\%$ and $5\%$, isotropy is approximately
obtained at zero deformation; the glass has recovered its original
state. However, unloading from larger deformations
$\varepsilon_{max}=7.5\%$, $12.5\%$, $17.5\%$ and $25\%$ we observe
that isotropy is never fully recovered. The beginning of the
unloading stage shows a linear evolution somewhat comparable to the
first elastic loading, consistent with the fact that the incremental
strain during unloading is expected to be purely elastic.\revision{It
is however noteworthy that a significant plastic strain occurs during
this unloading phase showing a strong kinematic hardening, and a very
severe shrinkage of the elastic domain.} The permanent plastic shear
deformation obtained at zero shear stress can be associated to the
minimal gap between the eigenvalues of the fabric tensor but this gap
is clearly growing well beyond the numerical uncertainty.

The same analysis can be performed in a more quantitative manner by
following the evolution of the scalar anisotropy index $\alpha$ under
these shear loading and unloading tests and is presented in
Fig. \ref{fig3}.  The anisotropy index decreases under unloading,
reaches a minimum value, $\alpha_{min}$, approximately when the shear
stress is null.  If the load is reversed to negative values, $\alpha$
rises again to saturate at a similar level as for direct shear.  For a
maximum deformation $\varepsilon_{max}=2.5\%$, $5\%$, we recover after
unloading $\alpha_{min} \approx 10^{-3}$: no change is noticeable when
compared with the original state. For larger deformations, however, a
clear increase of $\alpha_{min}$ is observed up to values close to
$10^{-2}$, well above the numerical uncertainty.

\begin{figure}[t]
\begin{center}
\epsfig{file=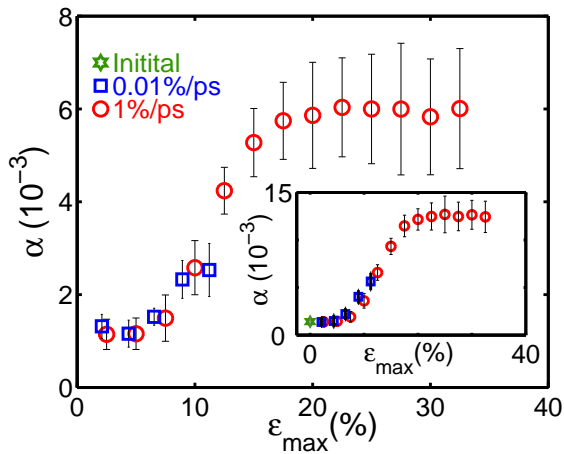,width=0.45\textwidth}
\caption{ Anisotropy parameter after unloading $\alpha_{min}$ vs.
maximum strain $\varepsilon_{max}$ for two different imposed strain
rates 10$^{-2}$ ps$^{-1}$ (black circles) and 10$^{-4}$ ps$^{-1}$ (red
squares). After shear unloading, normal stresses persist due to the
plane deformation geometry and amplify the structural anisotropy (see
insert). After relaxation of these stresses with an additional NPT
step of 20 ps, a stress free state is obtained while the silica
structure still exhibits a significant anisotropy index.
\label{fig4}}
\end{center}
\end{figure}

These results are summarized in Fig.~\ref{fig4} which shows the
evolution of the minimum gap, $\alpha_{min}$, versus the maximum
deformation, $\varepsilon_{max}$.  The results obtained with a much
slower strain rate $10^{-4}{\mathrm ps}^{-1}$, (one hundred times
slower than the previous case) are also shown in the same figure. It
is remarkable that both curves almost superimpose each other. Once
silica has experienced shear plasticity, a significant anisotropy
persists even after unloading.  Note that \revision{an additional
relaxation towards zero stress state (NPT ensemble) during 20 ps
after unloading was performed. This ensures to eliminate normal
stresses due to the shearing in plane deformation. As shown in the
insert, the latter naturally induce an additional anisotropy of
trivial elastic origin which has to be eliminated before attesting
for the presence of the plasticity induced anisotropy discussed in
the present study.}

On the basis of molecular dynamics study of plasticity of amorphous
silica under simple shear, a non-reversible anisotropy sets in and
appears stable (at the time scale of MD).  Detailed structure
investigation could not reveal any significant deviation from the
amorphous structure based on {\it e.g.} interatomic distances The
eigenvectors of these Si-O-Si directions are aligned with the
principal strain/stress directions.  Because of the relatively short
time scale naturally associated with MD simulations, an experimental
validation of this observation is needed to validate our conclusions,
in particular concerning the stability of this phase. Experimentally,
anisotropy is expected to give rise to birefringence, a property
which might be more easily accessible than structural analyses. Using
standard photoelasticity parameters for silica and the maximum
anisotropy obtained in the present work $\alpha=5 10^{-3}$, a crude
estimate of the birefringence gives an index contrast $\Delta
n=10^{-3}$. Such a phenomenon may for instance be tested in the
framework of rotational anvil cell
experiments\cite{Levitas-JPCS06,Levitas-JPCM06}.



\noindent {\bf \revision{Acknowledgments}}
 We acknowledge D. Bonamy. R. Kalia, and L. Van
Brutzel. CLR wishes to acknowledge the National Science Foundation
Graduate Research Fellowship under grant no. 0401467 for financial
support. DV, SR and MT acknowledge the financial support of ANR grant
``Plastiglass'' no. ANR-05-BLAN-0367-01.

\bibliographystyle{apsrev}
\bibliography{depinning,vdb,plasticity,silica}

\end{document}